\documentclass[runningheads]{llncs}
\usepackage{graphicx}
\usepackage{float}
\usepackage{tikz}
\usetikzlibrary{shapes.geometric, arrows}
\usepackage{multirow}

\begin{document}
\title{Data Mining Approach to Analyze \\
Covid19 Dataset of Brazilian Patients}


\author{Josimar Edinson Chire Saire\inst{1}
\authorrunning{Chire Saire, J.}
\institute{Institute of Mathematics and Computer Science (ICMC) \\
University of Sao Paulo (USP)\\
Sao Carlos, SP, Brazil}\\
\email{jecs89@usp.br}}

\maketitle              
\begin{abstract}

The pandemic originated by coronavirus(covid-19), name coined by World Health Organization during the first month in 2020. Actually, almost all the countries presented covid19 positive cases and governments are choosing different health policies to stop the infection and many research groups are working on patients data to understand the virus, at the same time scientists are looking for a vacuum to enhance imnulogy system to tack covid19 virus. One of top countries with more infections is Brazil, until August 11 had a total of 3,112,393 cases. Research Foundation of Sao Paulo State(Fapesp) released a dataset, it was an innovative in collaboration with hospitals(Einstein, Sirio-Libanes), laboratory(Fleury) and Sao Paulo University to foster reseach on this
trend topic. The present paper presents an exploratory analysis of the datasets, using a Data Mining Approach, and some inconsistencies are
found, i.e. NaN values, null references values for analytes, outliers on results of analytes, encoding issues. The results were cleaned datasets for future studies, but at least a 20\% of data were discarded because of non numerical, null values and numbers out of reference range.
\end{abstract}

\section{Introduction}
The outbreak of Coronavirus(Covid19) started with first cases on December 2019, in Wuhan(China). The first reported case\cite{first_case} in South America was in Brazil on 26 February 2020, in São Paulo city. The strategy to stop the infections in the country was a partial lockdown to avoid the propagation of the virus.

On 28 January 2020, Ministry of Health of Brazil reported a suspected case of Covid19 in Belo Horizonte, Minas Gerais state, recently one student returned from China  \cite{first_case_1}, \cite{first_case_2}. The same day were reported two suspected cases  in Porto Alegre and Curitiba \cite{first_case_3}. 
The first confirmed COVID-19 case \cite{first_case_4} were reported in Brazil, a man of 61-year-old who returned from Italy. The patient was tested in Israelita Einstein Hospital in Sao Paulo state. On 14 May\cite{peak_cases}, more than 200 000 cases were confirmmed, this number double during the first days of May. 

Until August 11, the numbers of Brazil\footnote{Data extracted from website: https://virusncov.com/} are: total of 3,112,393 cases, with an increasing rate of new cases of 44,255(+1.4\%) and a total of 2,243,124 recovered cases.

Nowdays, many scientists are working around coronavirus covid19, but searching for conducted studies in South America, there is only a few number. After a searching in IEEX Xplorer using coronavirus, covid19 terms, one paper with Brazilian Affiliation is found \cite{paperbr_1}, related to data augmentation for covid19 detection.  Considering a preprint repository related to Medicine(Medxriv), using terms: covid19, coronavirus, data mining more than 50 papers are found. 


The table \ref{tab:top10} presents the top 10 results of MedxRiv query. Four of this papers is a conducted study for  South America countries and there is any work analyzing Brazilian context. In spite of, there is 4 papers with Brazilian Affiliation.

\begin{table}[hbpt]
\label{tab:top10}
\centering
\resizebox{\textwidth}{!}{
\tiny

\begin{tabular}{|l|c|c|c|c|}
\hline
\textbf{Author}      & \textbf{Title}                                                                                                                                                                                                         & \textbf{\begin{tabular}[c]{@{}c@{}}Country \\ \\ of Study\end{tabular}}       & \textbf{Keywords}                                                                                                                     & \textbf{Affiliation}                                                                                                                                                                                      \\ \hline
\cite{paper1} & \begin{tabular}[c]{@{}c@{}}Covid19 Surveillance in Peru \\ on April using Text Mining\end{tabular}                                                                                                                     & Peru                                                                          & \begin{tabular}[c]{@{}c@{}}Natural Language Processing, Text Mining,\\ People behaviour, Coronavirus, Covid-19\end{tabular}           & \begin{tabular}[c]{@{}c@{}}University of Sao Paulo(Brazil), \\ Universidad Privada del Norte(Peru)\end{tabular}                                                                                           \\ \hline
\cite{paper2} & \begin{tabular}[c]{@{}c@{}}Text Mining Approach\\  to Analyze Coronavirus\\  Impact: Mexico City as Case of Study\end{tabular}                                                                                         & Mexico                                                                        & \begin{tabular}[c]{@{}c@{}}Natural   Language   Processing,   Text   Mining,\\ People behaviour, Coronavirus, Covid-19\end{tabular}   & \begin{tabular}[c]{@{}c@{}}University of Sao Paulo(Brazil),\\ Tecnologico Nacional del Mexico /\\ Instituto Tecnologico de Matamoros \\ (Mexico)\end{tabular}                                                 \\ \hline
\cite{paper3} & \begin{tabular}[c]{@{}c@{}}How was the Mental Health of \\ Colombian people  on March \\ during Pandemics Covid19?\end{tabular}                                                                                  & Colombia                                                                      & Not available                                                                                                                         & University of Sao Paulo(Brazil),                                                                                                                                                                          \\ \hline
\cite{paper4} & \begin{tabular}[c]{@{}c@{}}Mining Twitter Data on\\ COVID-19 for Sentiment analysis\\  and frequent patterns Discovery\end{tabular}                                                                                    & Algiers                                                                       & \begin{tabular}[c]{@{}c@{}}tweets Analytics, COVID-19, sentiment \\ analysis, frequent patterns, association \\ rules mining\end{tabular} & \begin{tabular}[c]{@{}c@{}}University of Science and \\  Technology Houari Boumediène \\ (Algiers)\end{tabular}                                                                                             \\ \hline
\cite{paper5} & \begin{tabular}[c]{@{}c@{}}Infoveillance based on \\ Social Sensors to Analyze\\  the impact of Covid19 \\ in South American Population\end{tabular}                                                                   & \begin{tabular}[c]{@{}c@{}}South \\ America\\ (not Brazil)\end{tabular} & Not available                                                                                                                         & University of Sao Paulo(Brazil),                                                                                                                                                                          \\ \hline
\cite{paper6} & \begin{tabular}[c]{@{}c@{}}Spread of SARS-CoV-2 Coronavirus \\ likely constrained by climate\end{tabular}                                                                                                              & \begin{tabular}[c]{@{}c@{}}Not \\ \\ applicable\end{tabular}                  & Not available                                                                                                                         & \begin{tabular}[c]{@{}c@{}}National Museum of Natural \\ Sciences (Spain), \\ University of Évora (Portugal), \\ University of Helsinki (Finland)\end{tabular}                                                \\ \hline
\cite{paper7} & \begin{tabular}[c]{@{}c@{}}The Role of Host Genetic Factors\\ in Coronavirus  Susceptibility: \\ Review of Animal and \\ Systematic Review of Human Literature\end{tabular}                                            & \begin{tabular}[c]{@{}c@{}}Not \\ \\ applicable\end{tabular}                  & \begin{tabular}[c]{@{}c@{}}Coronavirus; COVID-19; \\ Host  genetic factors ; SARS-CoV-2\end{tabular}                                  & \begin{tabular}[c]{@{}c@{}}University of Florida College of \\ Veterinary Medicine(Usa), \\ National Institutes of Health(Usa), \\ Johns Hopkins Bloomberg School \\ of Public Health ,(Usa)\end{tabular} \\ \hline
\cite{paper8} & \begin{tabular}[c]{@{}c@{}}Early epidemiological assessment \\ of the transmission potential\\ and virulence of coronavirus \\ disease 2019 (COVID-19) \\ in Wuhan City: China, \\ January-February, 2020\end{tabular} & China                                                                         & Not available                                                                                                                         & \begin{tabular}[c]{@{}c@{}}University Yoshida(Japan), \\ Kyoto University(Japan),\\ Georgia State University(Usa)\end{tabular}                                                                            \\ \hline
\cite{paper9} & \begin{tabular}[c]{@{}c@{}}Analysis of Epidemic Situation of \\ New Coronavirus Infection at Home \\ and Abroad Based \\ on Rescaled Range (R/S) Method\end{tabular}                                                   & China                                                                         & Not available                                                                                                                         & 
\begin{tabular}[c]{@{}c@{}} Sichuan Academy of Social Sciences \\ (China) \end{tabular}
                                                                                                                                                                 \\ \hline
\cite{paper10} & \begin{tabular}[c]{@{}c@{}}State heterogeneity of human mobility \\ and COVID-19 epidemics in \\ the European Union\end{tabular}                                                                                          & \begin{tabular}[c]{@{}c@{}}European \\ \\ Union\end{tabular}                  & \begin{tabular}[c]{@{}c@{}}Coronavirus 2019, epidemics, geographic, \\ trends, public health intervention\end{tabular}                & \begin{tabular}[c]{@{}c@{}}Shanghai Jiao Tong University \\  School of Medicine(China),\\ University at Buffalo(Usa), \\ Yale University School of Medicine(Usa)\end{tabular}                        \\ \hline
\end{tabular}

}
\caption{Ten results of Medrxiv Query about covid19 papers in South America} 
\end{table} 



Considering, the previous evidence it is necessary to conduct studies with Brazilian data, then the initiative of Fapesp is valuable to foster research on covid19 topic. 
The actual paper uses Data Mining Approach to perform an exploratory analysis of the dataset of Brazilian patients of Sao Paulo State. The methodology to explore data is presented in Section \ref{sec:2}, the experiments and results in Section \ref{sec:3}. Conclusion states in Section \ref{sec:4}, final recommendations and future work are presenten in Section \ref{sec:5}, \ref{sec:6}.

\section{Methodology}
\label{sec:2}

\begin{figure}[hbpt]
\centerline{\includegraphics [width=0.9\textwidth]{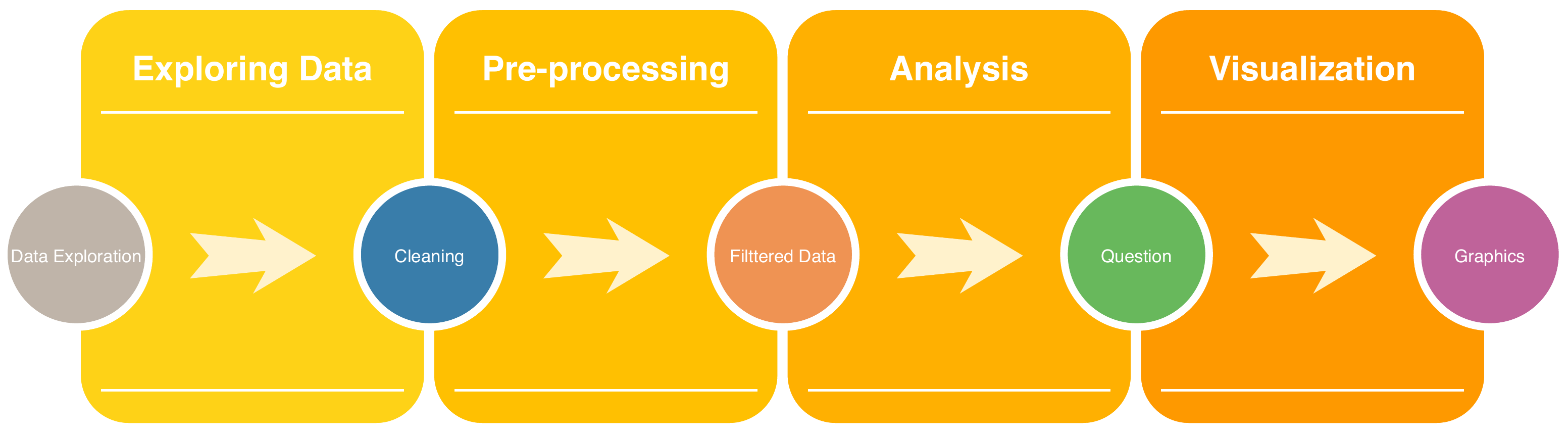}}
\caption{Methodology }
\label{fig:process}
\end{figure}

The conducted work follows a methodology inspired in CRISP-DM\cite{crisp2000}. The image \ref{fig:process} presents the flow between the phases of the exploration.

\subsection{Exploring data}

This step involves: check format files, open the files using a Language Programming or a tool. Review number of registers or rows per each file. Check existence of null values, check kind of each variable or field. For this step, Python Language Programming and pandas package are used to manipulate the data.

\subsection{Pre-processing Data}

This step is related how to deal with data before of generate graphics for analysis.

\begin{itemize}
\item If a specific variable must be numerical, but there is string values, so it is discarded
\item If null values are found, a discarding process must be considered.
\item If range reference for one exam, analytes is null then the analysis is not possible.
\end{itemize}

\subsection{Analysis}

Using clean data is possible to answer some questions related to age distribution, sex distribution, distribution of results to detect anomalies or outliers. The questions can require a kind of specific graphic to suppot analysis. 

\subsection{Visualization}

Considering distribution of few classes, a pie chart is useful to check proportions, subsection \ref{subsec:sex}, \ref{subsec:covid19} . For age distribution, bar plot can show how is the distribution, see subsection \ref{subsec:age}, \ref{subsec:date}, \ref{subsec:freq}. The analysis is dozen of values can be supported for boxplot graphics, in subsection \ref{subsec:box1}, \ref{subsec:box2}.

\clearpage

\section{Experiments and Results}
\label{sec:3}

\subsection{Datasets}

The release of the datasets is the result of collaboration between Research Foundation (FAPESP)\cite{covid_dataset}, Fleury Institute, Israelita Albert Einstein Hospital, Sirio-Libanes Hospital and the University of Sao Paulo. The goal is to contribute and promote research related to Covid19. The datasets share the data dictionaries of Patients(see Tab. 1), Test (Tab. 2).

\begin{table}[hbpt]
\label{tab:patient}
\caption{Data Dictionary of Patient Dataset- Einstein, Fleury, Sirio-Libanes Hospital}
\centering
\resizebox{\textwidth}{!}{
\tiny
\begin{tabular}{|l|l|l|l|}
\hline
\textbf{Variable} & \textbf{Description} & \textbf{Format} & \textbf{Content}                                                               \\ \hline
                                         &                                             &                                        &                                                                                                                                                                                                                                                                                                                                                                                                                                                                                                                 \\ \hline
ID\_PACIENTE                             & Unique identification of patient            & Alphanumeric characters                & String, key patient                                                                                                                                                                                                                                                                                                                                                                                                                                                                                             \\ \hline
IC\_SEXO                                 & Genre                                       & Alphanumeric character                 & \begin{tabular}[c]{@{}l@{}}F - Feminino(Female)\\ M - Masculino(Male)\end{tabular}                                                                                                                                                                                                                                                                                                                                                                                                                              \\ \hline
AA\_NASCIMENTO                           & Birth date                                  & Number                     & \begin{tabular}[c]{@{}l@{}}Example: 1959\\ (*) AAAA - for people was born before or equel 1930\end{tabular}                                                                                                                                                                                                                                                                                                                                                                                                     \\ \hline
CD\_PAIS                                 & Country of residence                        & Alphanumeric                           & Exemplo: BR                                                                                                                                                                                                                                                                                                                                                                                                                                                                                                     \\ \hline
CD\_UF                                   & Federal State Identifier                    & Alphanumeric characters                & \begin{tabular}[c]{@{}l@{}}AC - Acre, AL - Alagoas, AM - Amazonas, AP - Amapa, BA - Bahia,\\ CE - Ceará, DF - Distrito Federal, ES - Espirito Santo, GO - Goiás,\\ MA - Maranhão, MG - Minas Gerais, MS - Mato Grosso do Sul,\\ MT - Mato Grosso, PA  - Pará, PB - Paraíba, PE - Pernambuco,\\ PI - Piauí, PR - Paraná, RJ - Rio de Janeiro, RN - Rio Grande do Norte,\\ RO  - Rondônia, RR - Roraima, RS - Rio Grande do Sul,\\ SC - Santa Catarina, SE - Sergipe, SP - São Paulo, TO - Tocantins\end{tabular} \\ \hline
CD\_MUNICIPIO                            & Residence City & Alphanumeric                           & \begin{tabular}[c]{@{}l@{}}Example: SAO PAULO, CAMPINAS, SANTO ANDRE\\ MMMM - for the lowest occurrences \end{tabular}                                                                                          \\ \hline
CD\_CEP                                  & Postal Code                                 & Number (**)                       & First five digits of Postal Code, (**) CCCC - for low number of ocurrences                                                                                                                                                                                                                                                                                                                                                                                                                                            \\ \hline
\end{tabular}
}
\end{table}

\begin{table}[hbpt]
\label{tab:tests}
\caption{Data Dictionary of Tests - Einstein, Fleury, Sirio-Libanes Hospital}
\centering
\resizebox{\textwidth}{!}{
\tiny
\begin{tabular}{|l|l|l|l|}
\hline
\textbf{Variable Name} & \textbf{Description}                                                                       & \textbf{Format}                                                       & \textbf{Content}                                                                                                                                                                                                                                          \\ \hline
\textbf{}              & \textbf{}                                                                                  & \textbf{}                                                             & \textbf{}                                                                                                                                                                                                                                                 \\ \hline
ID\_PACIENTE           & Unique identification of patient                                                           & \begin{tabular}[c]{@{}l@{}}Alphanumeric\\ character\end{tabular}      & String, patient key                                                                                                                                                                                                                                       \\ \hline
DT\_COLETA             & Exam collection date                                                                             & Date (yyyy/MM/dd)                                                     & Date                                                                                                                                                                                                                                                      \\ \hline
DE\_ORIGEM             & Origin of patient   & \begin{tabular}[c]{@{}l@{}}Alphanumeric \\ character (4)\end{tabular} & \begin{tabular}[c]{@{}l@{}}HOSP – Exam made in a hospital\end{tabular}                                                                                                                                           \\ \hline
DE\_EXAME              & Description of Exam                                                          & Alphanumeric                                                          & Example: HEMOGRAMA(blood count)                                                                    \\ \hline
DE\_ANALITO            & Analyte description                                                                       & Alphanumeric                                                          & \begin{tabular}[c]{@{}l@{}}Example: Eritrócitos(Erythrocytes), \\ Leucócitos(Leukocytes), Glicose(Glucose) \\ 
\end{tabular}                                                            \\ \hline
DE\_RESULTADO          & \begin{tabular}[c]{@{}l@{}}Result of exam, \\ related to DE\_ANALITO\end{tabular} & Alphanumeric                                                          & \begin{tabular}[c]{@{}l@{}}If DE\_ANALITO requires numerical values, \\ Integer ou Float\\ If DE\_ANALITO requeries qualitative, \\ String with restrict domain \end{tabular}                                                                                       \\ \hline
CD\_UNIDADE            & Unit of measurement                                                                      & Alphanumeric                                                          & \begin{tabular}[c]{@{}l@{}}String\\ Exemplo: g/dL (grams por deciliter)\end{tabular}                                                                                                                                                                      \\ \hline
DE\_VALOR\_REFERENCIA  & \begin{tabular}[c]{@{}l@{}}Reference values\\ for DE\_RESULTADO\end{tabular}      & Alphanumeric                                                          & \begin{tabular}[c]{@{}l@{}}String - Reference value for de\_analito in \\  the population  $Min_{Value}$, $Max_{Value}$\\ Não Detectado(Not detected)/Detectado(Detected) \\ Example for glucose: 75 to 99\\ Example for progesterone: until 89\end{tabular} \\ \hline
\end{tabular}
}
\end{table}

The size of dataset are presented in Table 3 for three data sources. SL Hospital provided a dataset about outcomes of the patients.

\begin{table}[hbpt]

\centering
\tiny
\caption{Features of Dataset}
\begin{tabular}{c|c|c|c|}
\cline{2-4}
                                              & \textbf{Einstein Hospital}                                          & \textbf{Fleury}                                                    & \textbf{SL Hospital}                                                \\ \hline
\multicolumn{1}{|c|}{\textbf{Patient(size)}}  & 43,562                                                               & 129,596                                                             & 2,731                                                                \\ \hline
\multicolumn{1}{|c|}{\textbf{Test(size)}}     & 1,853,695                                                             & 2,496,591                                                            & 371,357                                                              \\ \hline
\multicolumn{1}{|c|}{\textbf{Test(Dates)}}    & \begin{tabular}[c]{@{}c@{}}2020-01-01 \\ to 2020-06-24\end{tabular} & \begin{tabular}[c]{@{}c@{}}2019-11-01\\ to 2020-06-15\end{tabular} & \begin{tabular}[c]{@{}c@{}}2020-02-26\\ to 2020-06-27\end{tabular}  \\ \hline
\multicolumn{1}{|c|}{\textbf{Outcome(size)}}  & -                                                                   & -                                                                  & 9,634                                                                \\ \hline
\multicolumn{1}{|c|}{\textbf{Outcome(Dates)}} & -                                                                   & -                                                                  & \begin{tabular}[c]{@{}c@{}}2020-02-26 \\ to 2020-06-29\end{tabular} \\ \hline
\end{tabular}
\end{table}

\subsection{Exploration}

This subsection present some graphics to describe data and let posterior analysis, besides the requeriment of some graphics related to distribution, i.e. bar plot, boxplot.

\subsubsection{Description of datasets}

\begin{figure}[hbpt]
\centerline{
\includegraphics [width=0.4\textwidth]{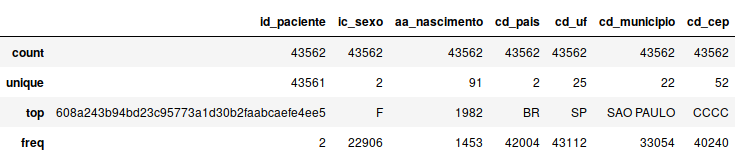}
\includegraphics [width=0.6\textwidth]{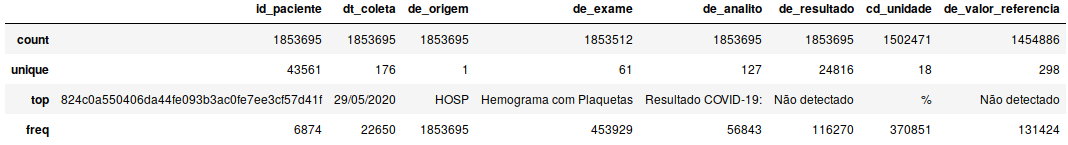}
}
\centerline{
\includegraphics [width=0.4\textwidth]{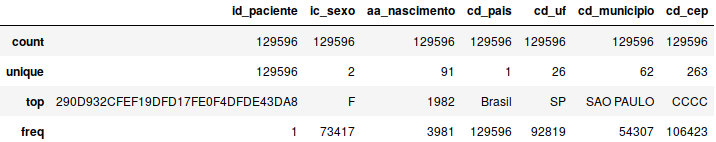}
\includegraphics [width=0.6\textwidth]{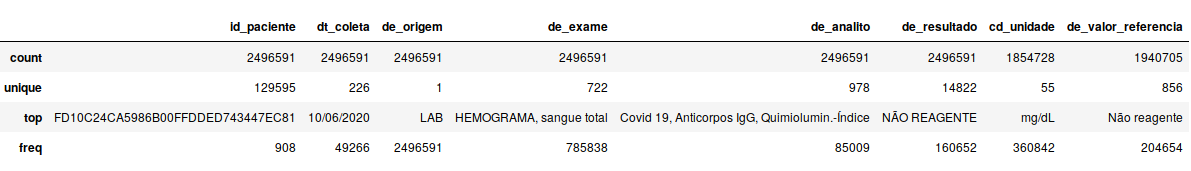}
}
\centerline{
\includegraphics [width=0.4\textwidth]{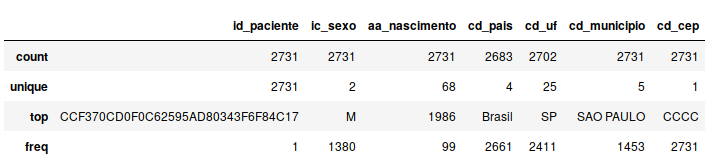}
\includegraphics [width=0.6\textwidth]{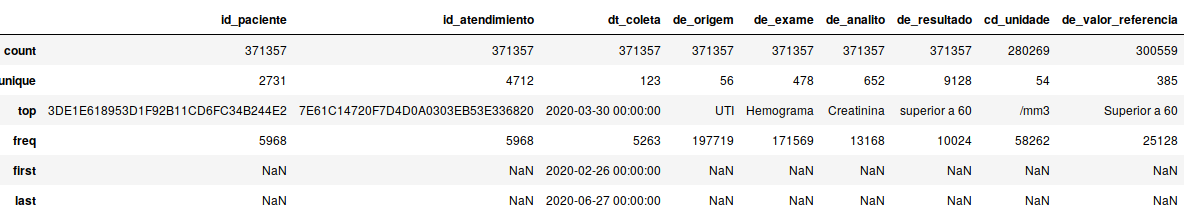}
}
\caption{(a) Einstein, (b) Fleury and (c) SL Datasets Description}
\label{fig:description}
\end{figure}

The Figure \ref{fig:description} is presented with counting values, unique values, top for each field. The name of columns were transformed to lowercase to have an uniform name of fields.

\begin{itemize}
    \item Figure 3.b presents a different number of id\_paciente in patient dataset and exam dataset, 129596(patient) 129595(exam).
    \item Einstein and SL Hospitals( cd\_pais ) presents people living in countries different than Brazil.
    \item The most frequent age of patients is: 38(Einstein, Fleury) and 34(SL).
    \item Female patients are higher in number in Einstein, Fleury.
    \item Most frequent cd\_uf, cd\_municipio is Sao Paulo State or city and CCCC is most common in Postal Code, so this places do not have meaningful number of ocurrences.
    \item Einstein and Fleury have a unique de\_origem: Hosp, Lab respectively. But SL Hospital has 56 different.
    \item The exam hemograma(blood count) is the most frequent in the datasets, and de\_analito more frequent in Eistein, Fleury are related to \textbf{Covid19}.
    \item Eistein has the lowest number of different de\_exame(61), de\_analito(127). Fleury has the highest de\_exame(722), de\_analito(978).  SL has de\_exame(478), de\_analito(652). Therefore, numer of de\_valor\_referencia are related.
    \item SL Hospital presentes NaN(Not a number) values, then it is possible find NaN values in the datasets.
\end{itemize}

\subsection{Sex Distribution}
\label{subsec:sex}

\begin{figure}[hbpt]
\centerline{
\includegraphics [width=0.25\textwidth]{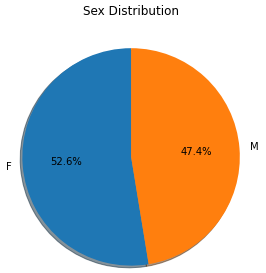}
\includegraphics [width=0.25\textwidth]{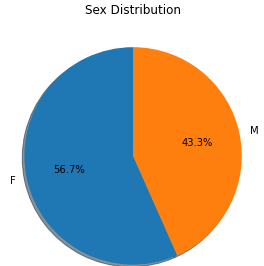}
\includegraphics [width=0.15\textwidth]{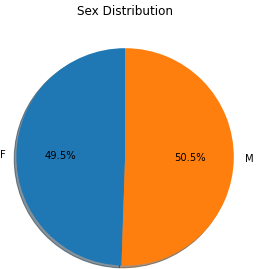}
}
\caption{Sex Distribution(Einstein, Fleury and HL) }
\label{fig:sex_dis}
\end{figure}

Female population is slightly bigger than male population in Einstein, Fleury but SL presents male population bigger for 0.05\%(29 people), see Fig. \ref{fig:sex_dis}.

\subsection{Age Distribution}
\label{subsec:age}

\begin{figure}[hbpt]
\centerline{
\includegraphics [width=0.45\textwidth]{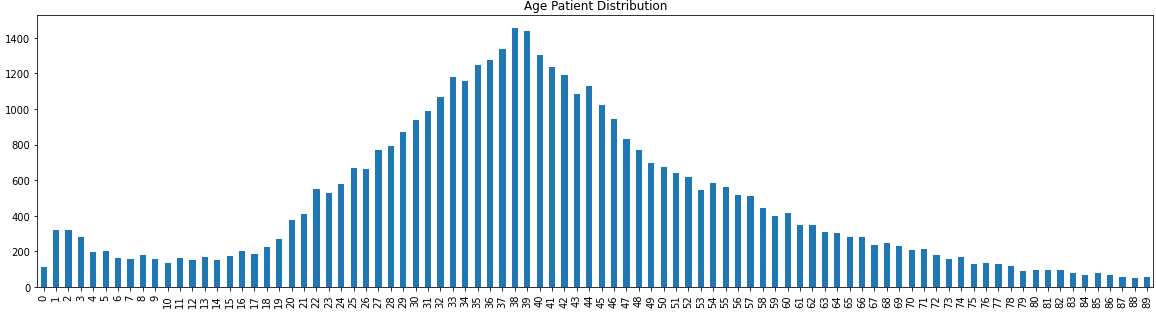}
\includegraphics [width=0.45\textwidth]{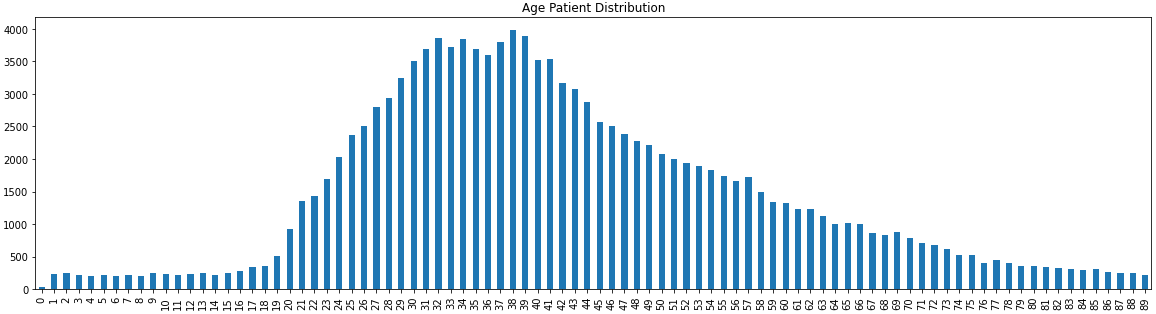}
}
\centerline{
\includegraphics [width=0.45\textwidth]{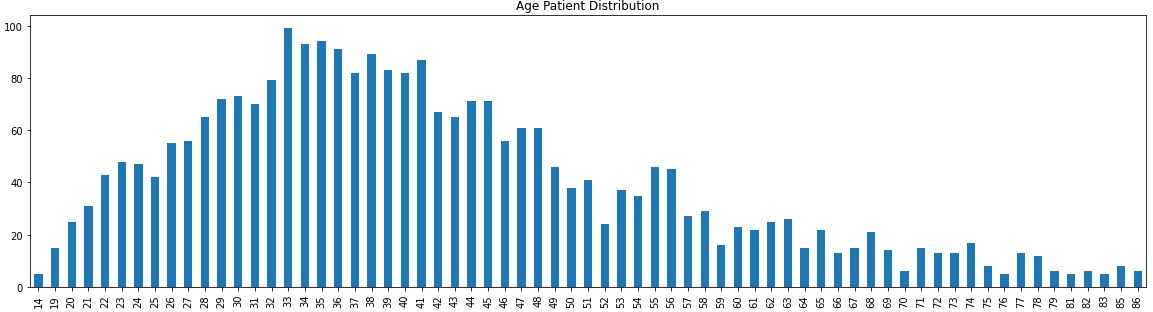}
}
\caption{Age Distribution (Einstein, Fleury, SL) }
\label{fig:age_dis}
\end{figure}

Datasets of Einstein, Fleury have younger patients from 0 to 14 until 89 but SL Hospital only from 14 to older(86), this graphics are presented in Fig. \ref{fig:age_dis}

\subsection{Date Collection of Exams}
\label{subsec:date}

\begin{figure}[hbpt]
\centerline{
\includegraphics [width=0.45\textwidth]{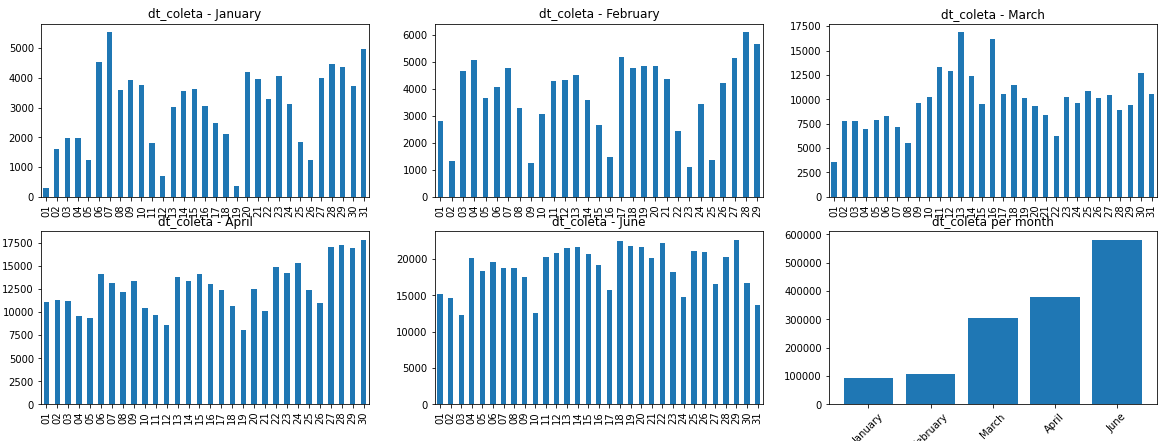}
\includegraphics [width=0.45\textwidth]{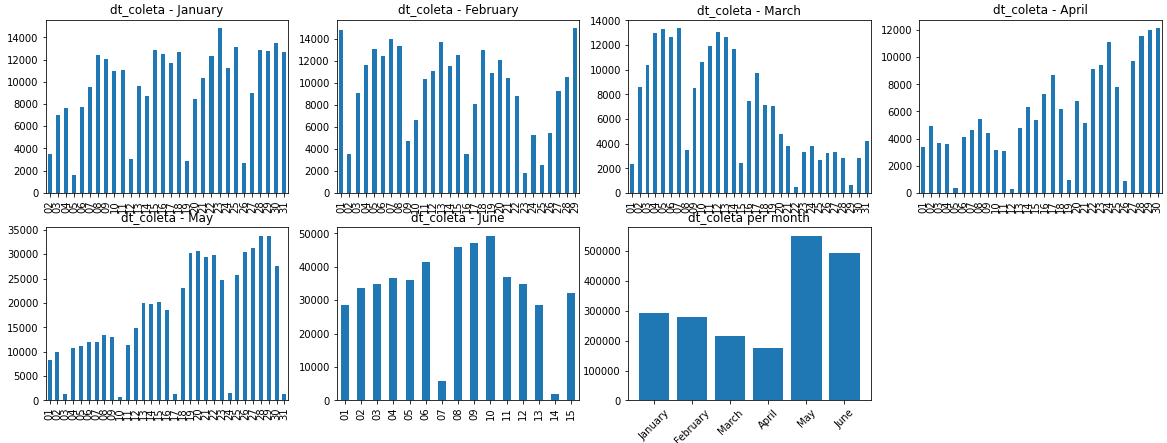}
}

\centerline{
\includegraphics [width=0.45\textwidth]{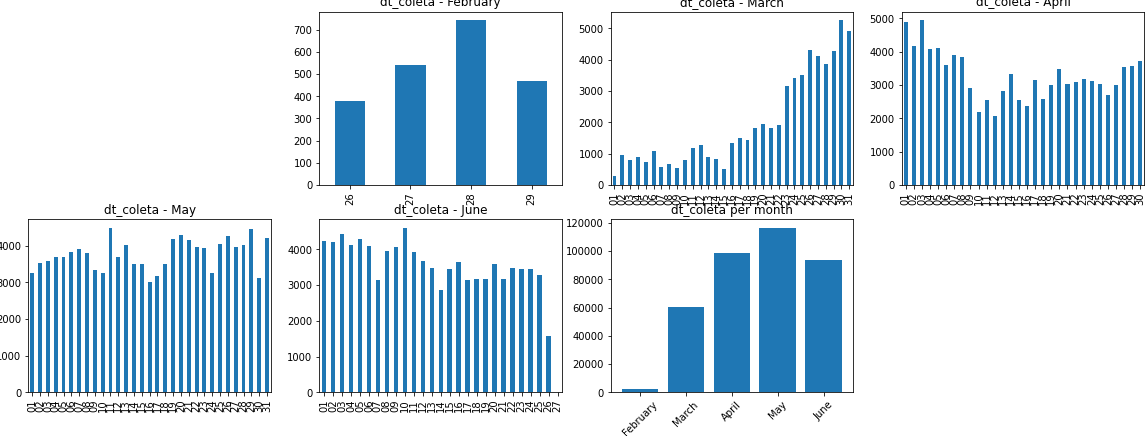}
}
\caption{Date Distribution (Einstein, Fleury, SL) }
\label{fig:date_dis}
\end{figure}

The graphic Fig. \ref{fig:date_dis} presents the number of collect exams per day and month, Einstein  presents an increasing number from January to June, Flury a decreasing from January to April but a peak on May, June. Besides, SL Hospital has an increasing from February to June.

\subsection{Most frequent exams per month}
\label{subsec:freq}

\begin{figure}[hbpt]
\centerline{
\includegraphics [width=0.45\textwidth]{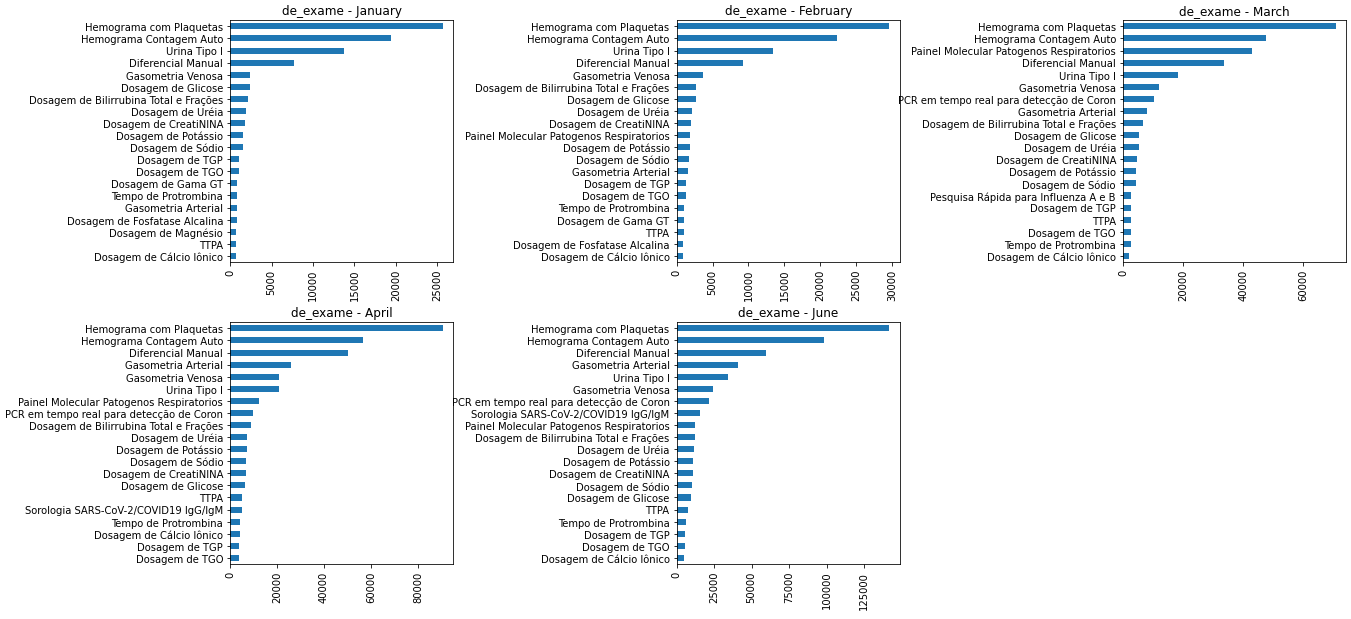}
\includegraphics [width=0.45\textwidth]{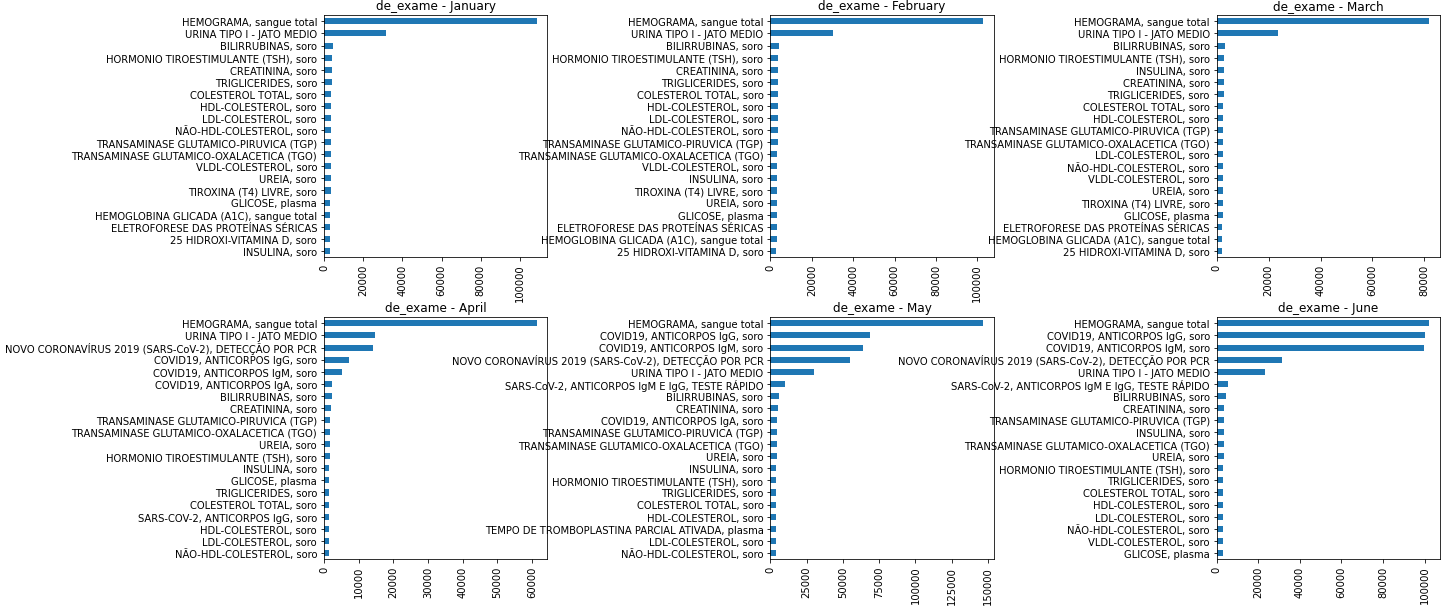}
}
\centerline{
\includegraphics [width=0.45\textwidth]{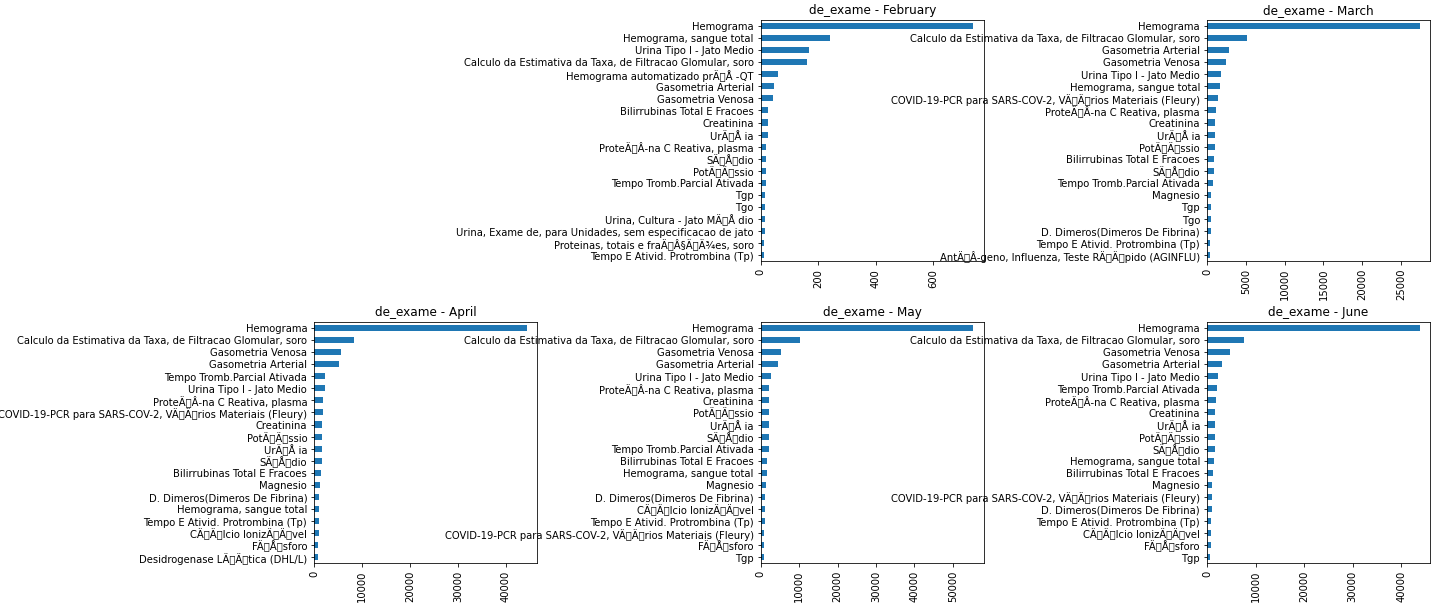}
}
\caption{Exam Distribution (Einstein, Fleury, SL) }
\label{fig:exams}
\end{figure}

To answer what were the most frequent exams during the month of each dataset, graphic Fig. \ref{fig:exams} presents the 20 most frequents.

\begin{itemize}
    \item Three datasets has blood count exam on the top of each month.
    \item Only Fleury has exams related to covid19 detection on April, May, June on the top 5.
    \item There are many kind of exams related to covid19 for Hospital, i.e. PCR, Sorologia SARS-Cov-2/Covid19 (Einstein). Fleury has NOVO Coronavirus 2019, Covid19 Anticorpos lgG, lgM, lgA and more. SL Hospital has Covid-19 PCR para Sars-Cov2 and a problem with encoding is detected in this dataset.
    \item For the previous reason, each dataset is studied separately.
\end{itemize}

\subsection{Most frequent analyte per month}

\begin{figure}[hbpt]
\centerline{
\includegraphics [width=0.45\textwidth]{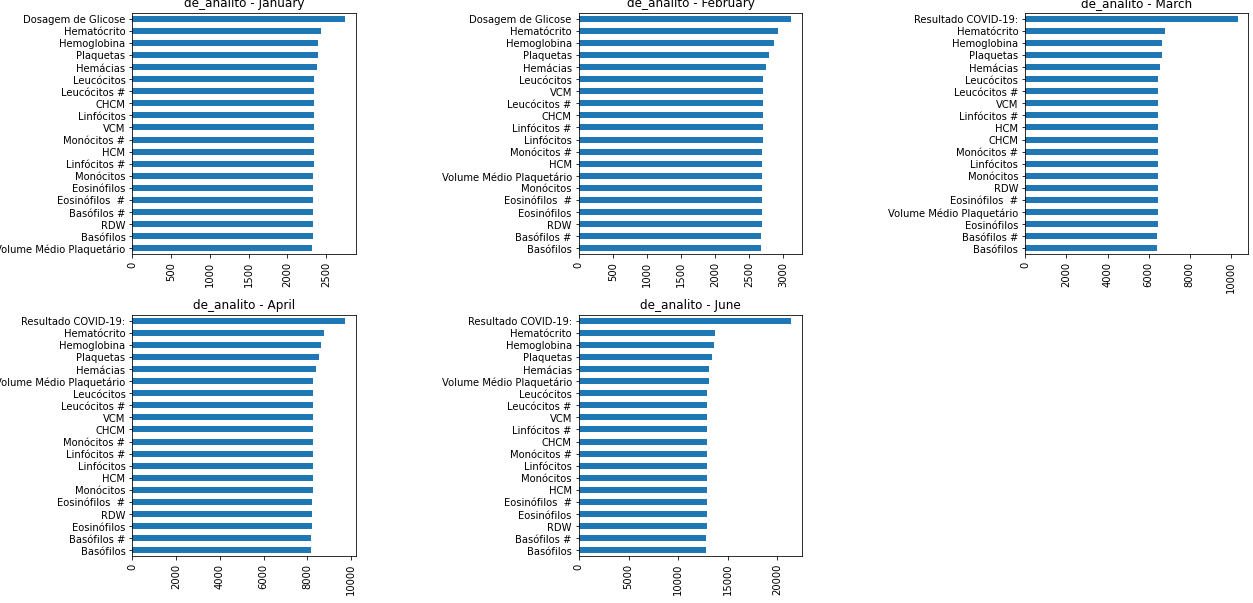}
\includegraphics [width=0.45\textwidth]{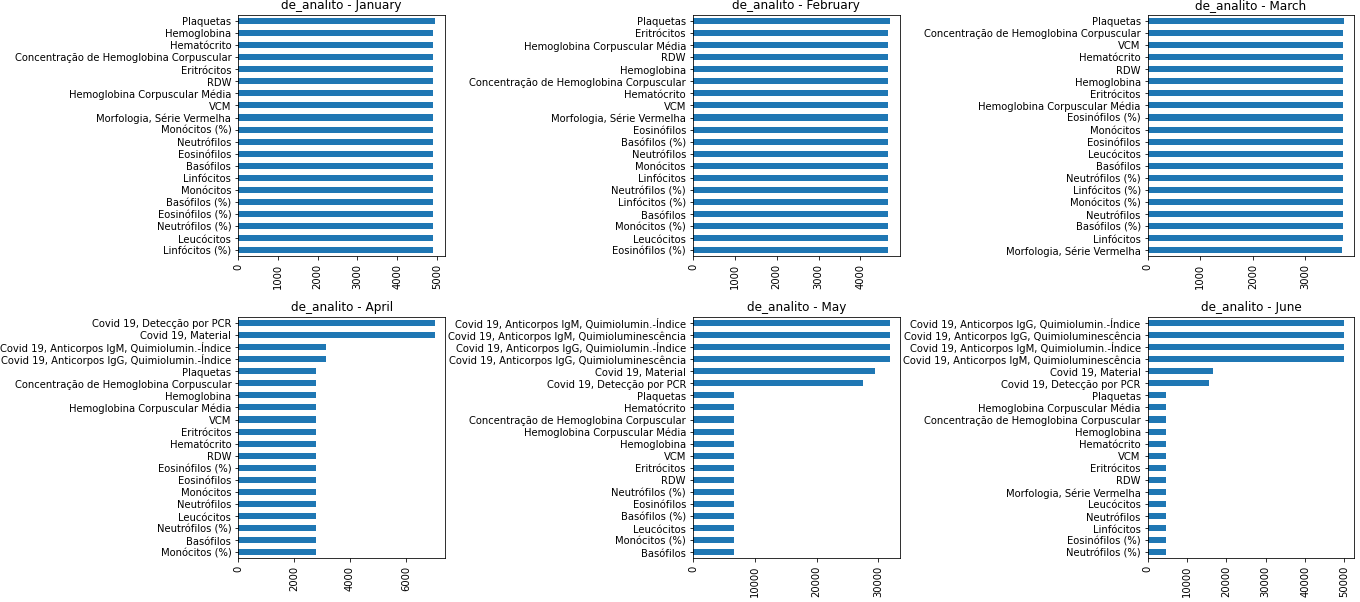}
}
\centerline{
\includegraphics [width=0.45\textwidth]{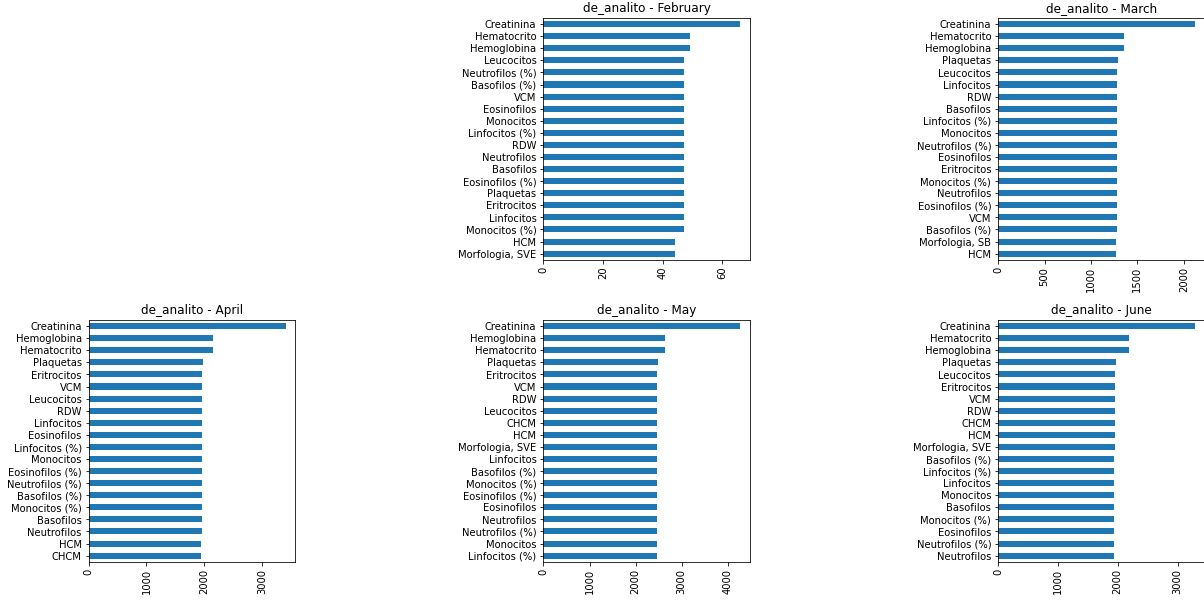}
}
\caption{Analyte per month(Einstein, Fleury, SL) }
\label{fig:exams1}
\end{figure}

Einstein and Fleury presents analytes related to covid19, i.e. resultado covid19, Covid19 deteccao por PCR, Covid19 material and more. Again, Fleury presents a variety of names for analytes related to covid19. And SL Hospital does not have any in the top 20(see Fig. \ref{fig:exams1}).

\subsection{Covid19 Analytes Distribution}
\label{subsec:covid19}

\begin{figure}[hbpt]
\centerline{
\includegraphics [width=0.65\textwidth]{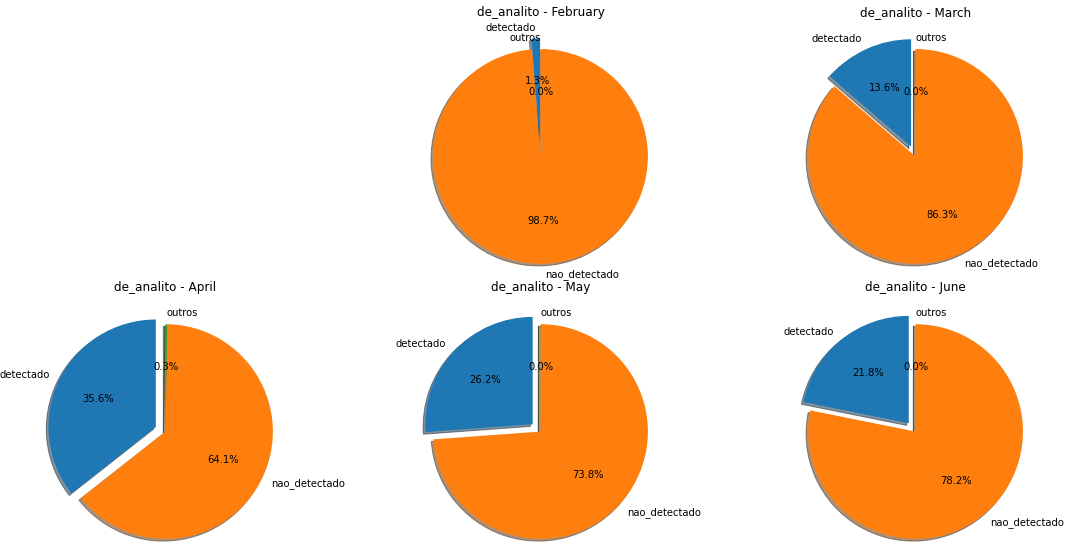}
}
\caption{Analyte per month(Einstein) }
\label{fig:acov}
\end{figure}

Considering analytes related to covid19, graphic \ref{fig:acov} presents the number of detected/not detected during the months for Hospital Einstein. Fleury and SL do not have an standardized outputs of covid19 exams, therefore is not possible to generate the graphics yet.

\subsection{Boxplot of most frequent exams}
\label{subsec:box1}

\begin{figure}[hbpt]
\centerline{\includegraphics [width=0.75\textwidth]{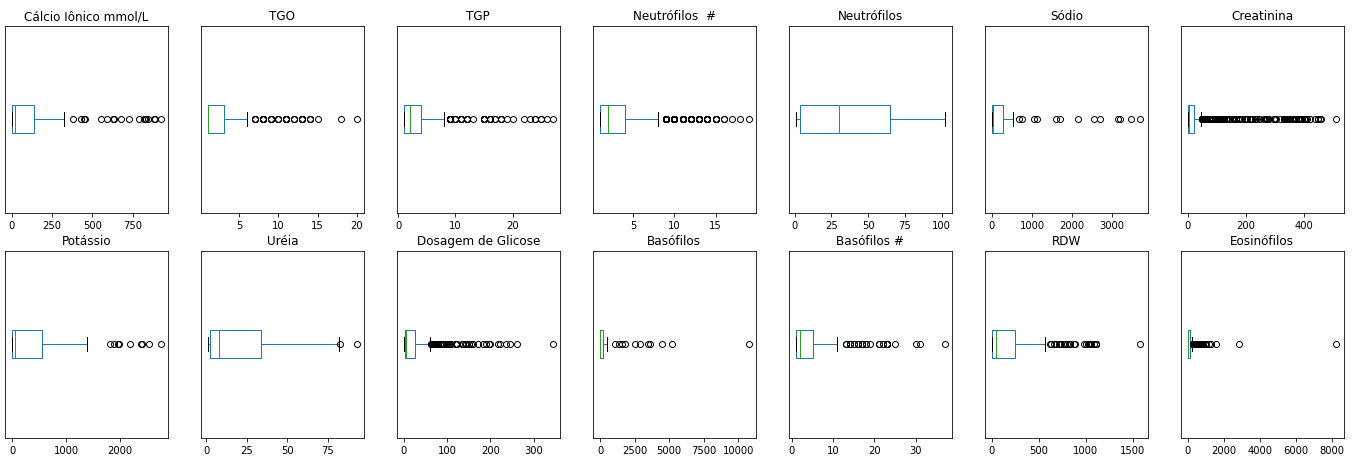}}
\caption{ Boxplot of top 14 analytes (Einstein) }
\label{fig:boxan}
\end{figure}

Considering top 14 of de\_analito and de\_resultado, the graphic Fig. \ref{fig:boxan} is presenting boxplot of the values of Einstein Hospital. It is necessary not to consider qualitative values, then only numerical values were used to build the plot. Analyzing the graphic is remarkable to many outliers in many of analytes, then a cleaning process is necessary.

\begin{figure}[hbpt]
\centerline{\includegraphics [width=0.75\textwidth]{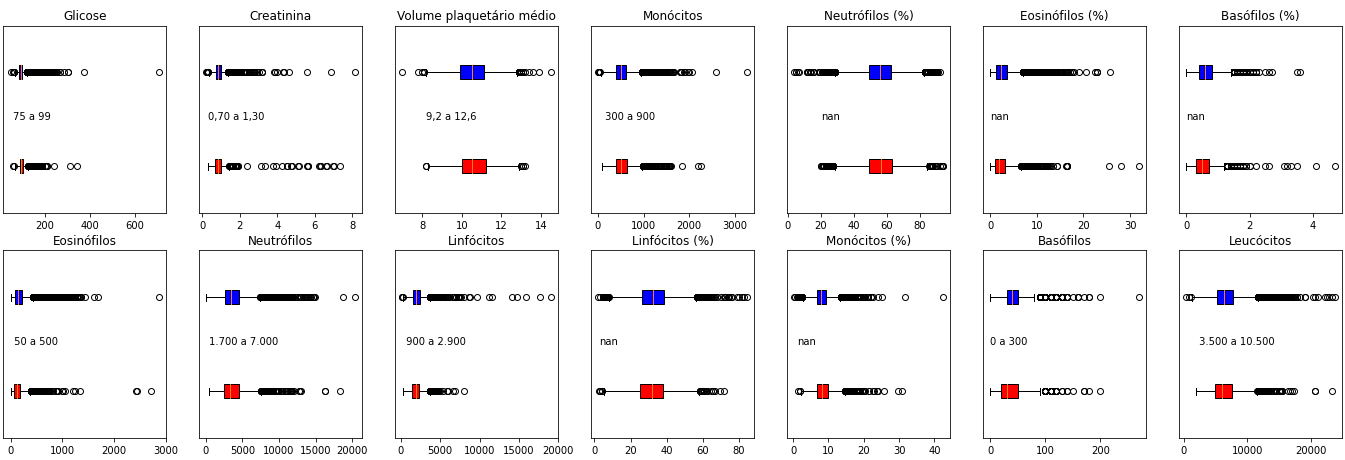}}
\caption{ Boxplot of top 14 analytes(Fleury) }
\label{fig:boxcov}
\end{figure}

Splitting data of covid19 detected and no detected, figure Fig. \ref{fig:boxcov} is presented. Again, outliers are present in Fleury dataset. Red ones(detected), blue(not detected).

\begin{figure}[hbpt]
\centerline{\includegraphics [width=0.7\textwidth]{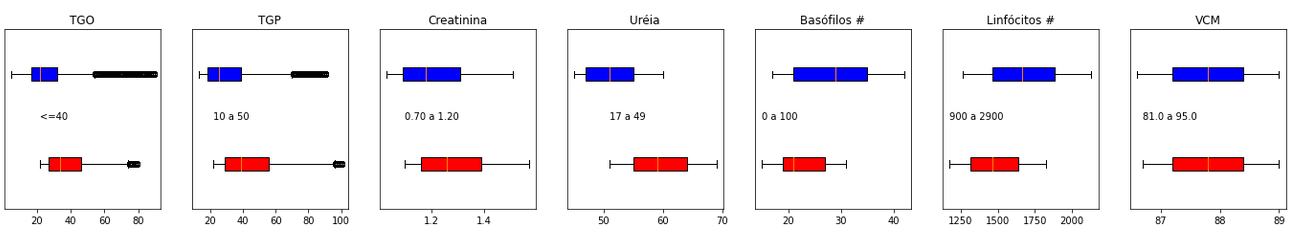}}
\centerline{\includegraphics [width=0.7\textwidth]{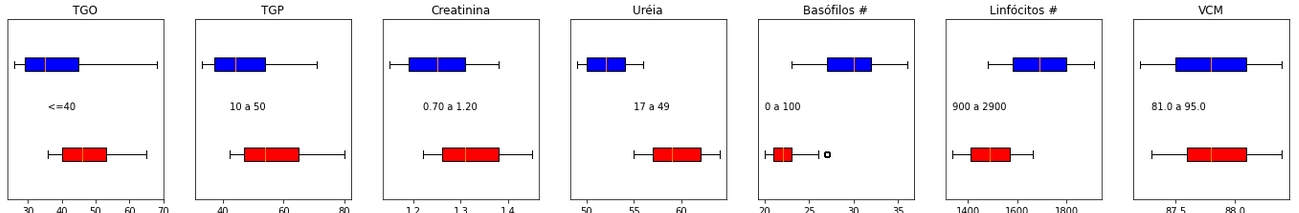}}
\caption{Boxplot of Cleaned dataset of Analytes with Abnormal Values, 0.5*std}
\label{fig:cbox1}
\end{figure}

Using a cleaning process using standard deviation(std) is proposed, because the outliers are further than median and in normal case two or three times higher is considered an abnormal value but in this situation, to have a better visualization of boxplot was used 0.5*std(see Fig. \ref{fig:cbox1}) and 0.2*std(see Fig. \ref{fig:cbox1}) on Einstein dataset considering analytes with abnormal values.

\subsection{Cleaning using Reference Values}
\label{subsec:box2}

The next graphics are created splitting Einstein dataset for genre. There is presence of NaN values in the reference value then these analytes are discared for the graphic, table \ref{tab:dis} presents the no valid de\_analito, it is a total of 8.

\begin{table}[hbpt]
\label{tab:dis}
\caption{No valid de\_analito for no valid reference range}
\centering
\begin{tabular}{|c|c|c|}
\hline
\textbf{De\_analito} & \textbf{Unity}            & \textbf{\begin{tabular}[c]{@{}c@{}}Range \\ \\ Reference\end{tabular}} \\ \hline
Neutrófilos          & \%                        & nan                                                                    \\ \hline
Dosagem de Glicose   & nan                        & nan                                                                    \\ \hline
Basófilos          & \%                        & nan                                                                    \\ \hline
Eosinófilos            & \%                        & nan                                                                    \\ \hline
Monócitos          & \%                        & nan                                                                    \\ \hline
Linfócitos           & \% & nan                                                                    \\ \hline
Leucócitos           & x10\textasciicircum{}3/uL & nan       

\\ \hline
Plaquetas           & x10\textasciicircum{}3/uL & nan       

\\ \hline
\end{tabular}
\end{table}

\begin{figure}[hbpt]
\centerline{\includegraphics [width=1.0\textwidth]{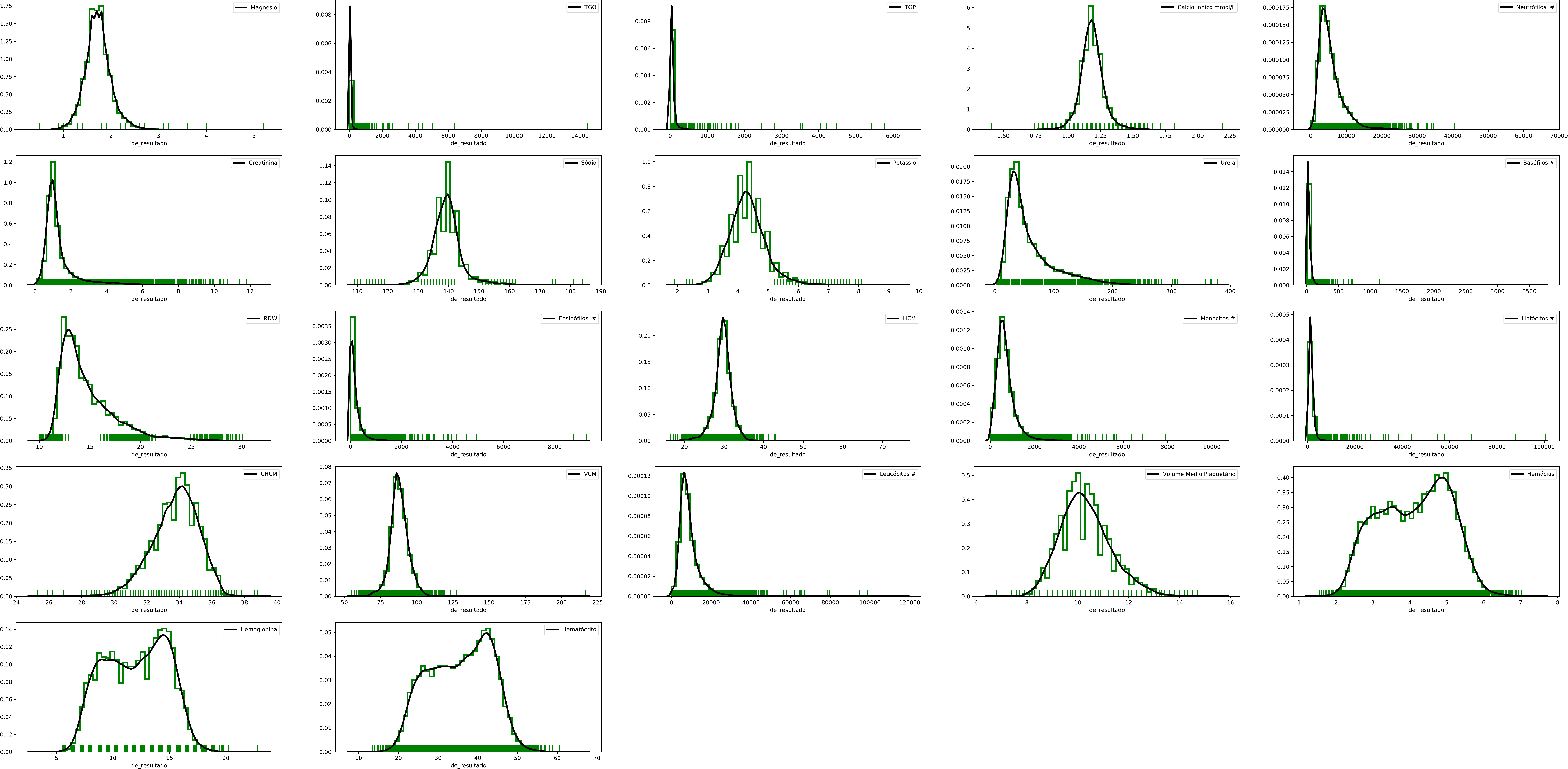}}
\caption{Men Analytes }
\label{fig:men_analitos}
\end{figure}

Plotting the distribution(Fig. \ref{fig:men_analitos}) for 30 most frequents analytes for men. 

\begin{figure}[hbpt]
\centerline{\includegraphics [width=1.0\textwidth]{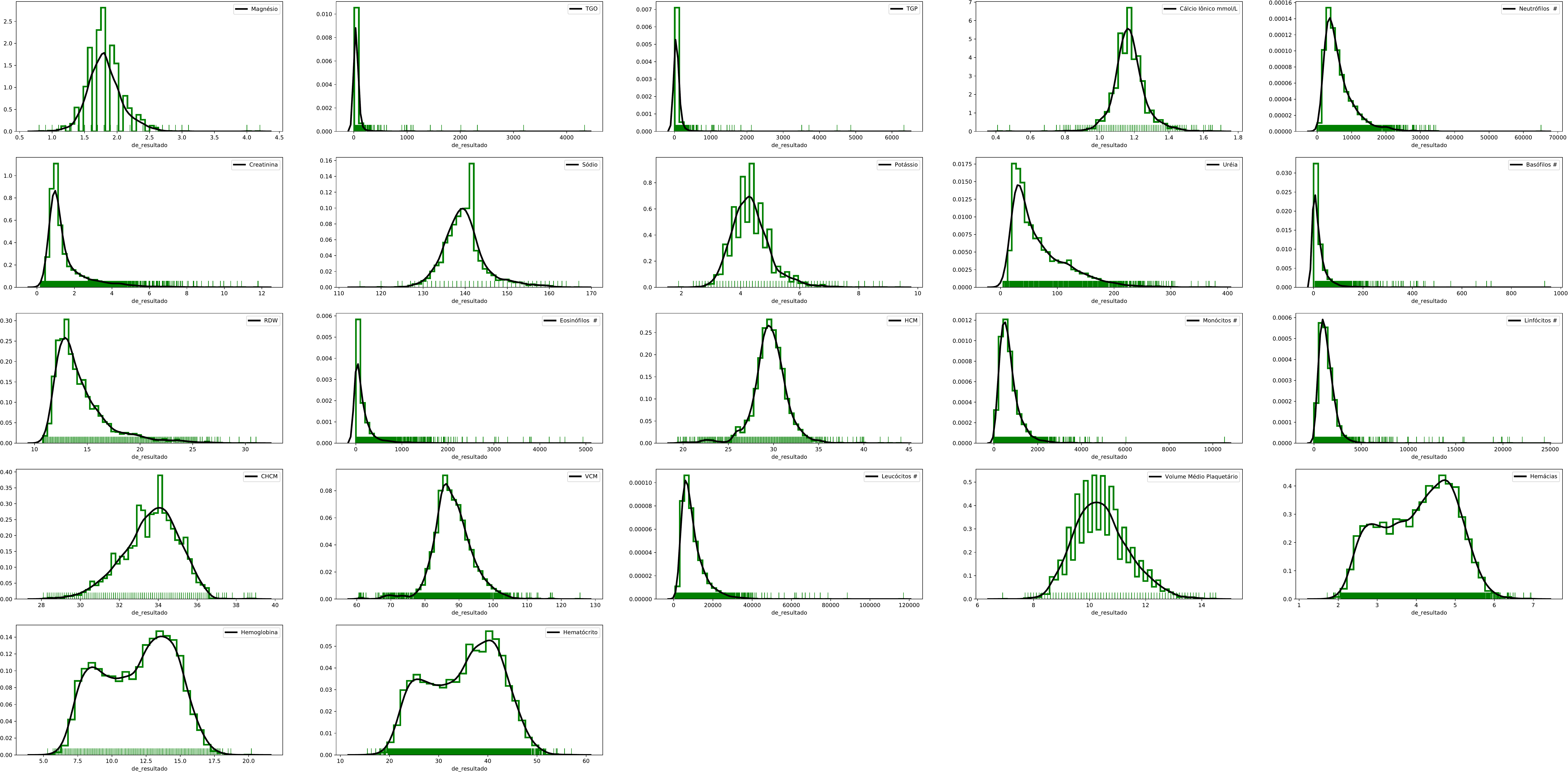}}
\caption{Men Analytes - Positives covid19 cases }
\label{fig:men_analitospos}
\end{figure}

The next graphic \ref{fig:men_analitospos} present the distribution for positive cases of covid19. In the two previous images \ref{fig:men_analitos} and \ref{fig:men_analitospos} is possible to observe a concentration of outliers in the sides of the normal distribution, i.e. TGO, TGP, Creatinina, Neutrófilos \#, Ureia. 

\begin{figure}[hbpt]
\centerline{\includegraphics [width=0.9\textwidth]{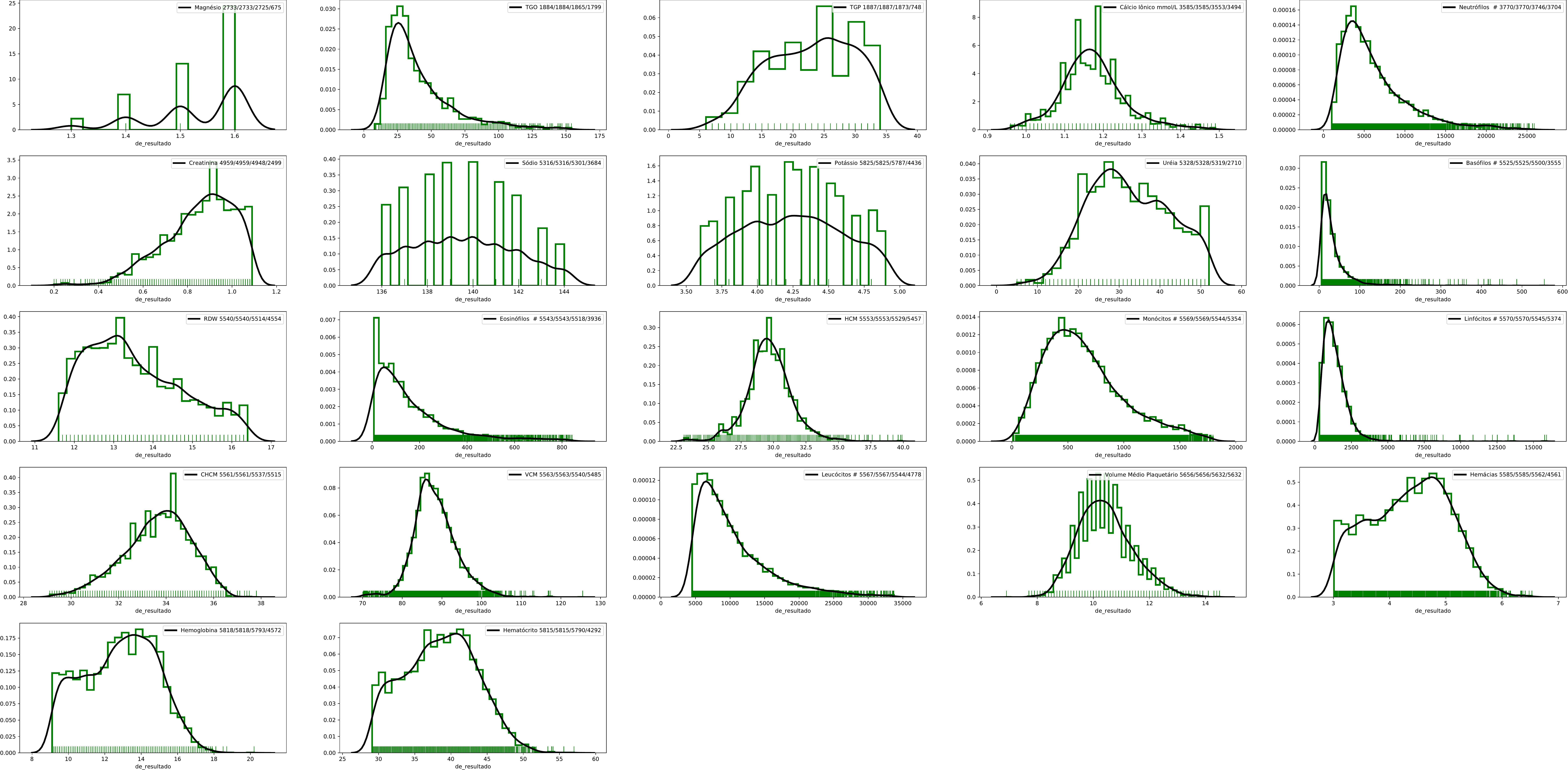}}
\caption{Filtered Men Analytes - Positives covid19 cases }
\label{fig:men_analitospos_fil}
\end{figure}

And graphic \ref{fig:men_analitospos_fil} introduces the result after of cleaning values and considering patients with positive cases and the date when it was detected until it finishes or open(no date for discard test). Because the aim of the analysis is understand how is the behaviour of the patients with positive diagnosis of covid19 during the active phase of virus, from the start until the end. Analyzing, Fig. \ref{fig:men_analitospos_fil}, it is possible to notice that the presence of outliers has disappeared, an exception with Basófilos \#.

Finally, Table \ref{tab:stat} presentss the steps used to clean data and generate Fig. \ref{fig:men_analitospos_fil}. First, only numerical values are considered, null values are discarded, and values out of reference range are not considered. For checking if values are inside of reference range, it was manually because there was many reference values too, only the lowest and highest value were used to filter data. Then, the reduction can be from 0.83 to 75.30 \%. An initial number of exams was 108,152 and final value after filtering 86,814 with a reduction of almost 20\% of the available data. Now, dataset is ready to answer more question and the research can continue.

\begin{table}[hbpt]
\label{tab:stat}
\caption{Reduction of Dataset}
\centering
\tiny
\begin{tabular}{|c|c|c|c|c|c|}
\hline
\textbf{de\_analito}     & \textbf{Initial} & \textbf{\begin{tabular}[c]{@{}c@{}}Only \\ \\ Numericals\end{tabular}} & \textbf{Not null} & \textbf{Range} & \textbf{Reduction} \\ \hline
Magnésio                 & 2733             & 2733                                                                   & 2725              & 675            & 75.30              \\ \hline
TGO                      & 1884             & 1884                                                                   & 1865              & 1799           & 4.51               \\ \hline
TGP                      & 1887             & 1887                                                                   & 1873              & 748            & 60.36              \\ \hline
Cálcio Iônico mmol/L     & 3585             & 3585                                                                   & 3553              & 3494           & 2.54               \\ \hline
Neutrófilos  \#          & 3770             & 3770                                                                   & 3746              & 3704           & 1.75               \\ \hline
Creatinina               & 4959             & 4959                                                                   & 4948              & 2499           & 49.61              \\ \hline
Sódio                    & 5316             & 5316                                                                   & 5301              & 3684           & 30.70              \\ \hline
Potássio                 & 5825             & 5825                                                                   & 5787              & 4436           & 23.85              \\ \hline
Uréia                    & 5328             & 5328                                                                   & 5319              & 2710           & 49.14              \\ \hline
Basófilos \#             & 5525             & 5525                                                                   & 5500              & 3555           & 35.66              \\ \hline
RDW                      & 5540             & 5540                                                                   & 5514              & 4554           & 17.80              \\ \hline
Eosinófilos  \#          & 5543             & 5543                                                                   & 5518              & 3936           & 28.99              \\ \hline
HCM                      & 5553             & 5553                                                                   & 5529              & 5457           & 1.73               \\ \hline
Monócitos \#             & 5569             & 5569                                                                   & 5544              & 5354           & 3.86               \\ \hline
Linfócitos \#            & 5570             & 5570                                                                   & 5545              & 5374           & 3.52               \\ \hline
CHCM                     & 5561             & 5561                                                                   & 5537              & 5515           & 0.83               \\ \hline
VCM                      & 5563             & 5563                                                                   & 5540              & 5485           & 1.40               \\ \hline
Leucócitos \#            & 5567             & 5567                                                                   & 5544              & 4778           & 14.17              \\ \hline
Volume Médio Plaquetário & 5656             & 5656                                                                   & 5632              & 5632           & 0.42               \\ \hline
Hemácias                 & 5585             & 5585                                                                   & 5562              & 4561           & 18.33              \\ \hline
Hemoglobina              & 5818             & 5818                                                                   & 5793              & 4572           & 21.42              \\ \hline
Hematócrito              & 5815             & 5815                                                                   & 5790              & 4292           & 26.19              \\ \hline
\textbf{Total}           & \textbf{108152}  &                                                                        &                   & \textbf{86814} & \textbf{19.73}     \\ \hline
\end{tabular}
\end{table}


\section{Conclusions}
\label{sec:4}
Coronavirus pandemic is active in the world, scientist are working to understand how to stop the virus, many areas are studying the covid19 impact in Heath, Economy therefore datasets related to patients are useful and important. Fapesp initiative to gather university and hospital is remarkable because it can foster research on the topic.

Real world datasets are not clean or ready for Data Mining or Data Science tasks then an exploratory phase is mandatory to see if data can be representative or useful to answer questions. Then, many cleaning steps were necessary to generate the final dataset and graphic, besides this cleaning step reduced the available dataset of men in 20\%, with a maximum value of 75.30\% for Magnesium Analyte, then it is possible a meanignful reduction of data is a cleaning task is performed. 

Finally, share the process of analysis is useful for researchers interested to analyze with this dataset, so it can save time, effort to future research.

\clearpage

\section{Recommendations}
\label{sec:5}
For researchers interested to work with these datasets, consider:
\begin{itemize}
    \item Check if range of dates for each dataset to know if this data is useful for your study.
    \item Sirio-Libanes Hospital has some issues related to encoding, this is the smallest dataset then you must analyze if it useful for analysis and search for the problems to fix them.
    \item Only Einsteing dataset has a standardized output for covid19 exams: detected or not detected. If you are from Computer Science or related field, this is better for your study. Because, Fleury has a variety of outputs, therefore is necessary the presence or advice of one person related to Medicine to explain you the different values.
    \item If you want to automatize filtering considering reference range of values, remember there are many for many analytes, then the suggestion is check this manually to check if it is possible to code the process.
\end{itemize}

\section{Future Work}
\label{sec:6}

For further work, a crossing of data is proposed to improve the analysis considering other variables, i.e. social-economic data, previous existence of health issues related to patients, considering data of other hospital to enhance the study.
By the other hand, a deep analysis will be performed with this new cleaned dataset.

\section*{Acknowledgement}

The author wants to thank to Fabio Faria, professor of UNIFESP(Federal University of São Paulo) for the invitation to analyze this dataset, to the team DS-Covid for the discussion about the generated graphics during the data analysis task, more news about future will be available in: https://dscovid.github.io/ .

%
%

\bibliographystyle{splncs04}
\bibliography{biblio.bib}

\begin{thebibliography}{10}
\providecommand{\url}[1]{\texttt{#1}}
\providecommand{\urlprefix}{URL }
\providecommand{\doi}[1]{https://doi.org/#1}

\bibitem{first_case_1}
Abril, E.: { Ministério da Saúde confirma 3 casos suspeitos de coronavírus
  no Brasil } (Jan 2020),
  \url{https://web.archive.org/web/20200129042253/https://exame.abril.com.br/brasil/ministerio-da-saude-confirma-3-casos-suspeitos-de-coronavirus-no-brasil/}

\bibitem{paper6}
Araujo, M.B., Naimi, B.: Spread of sars-cov-2 coronavirus likely to be
  constrained by climate. medRxiv  (2020). \doi{10.1101/2020.03.12.20034728},
  \url{https://www.medrxiv.org/content/early/2020/04/07/2020.03.12.20034728}

\bibitem{paper7}
Araujo, M.B., Naimi, B.: Spread of sars-cov-2 coronavirus likely to be
  constrained by climate. medRxiv  (2020). \doi{10.1101/2020.03.12.20034728},
  \url{https://www.medrxiv.org/content/early/2020/04/07/2020.03.12.20034728}

\bibitem{first_case}
AS/COA: { The Coronavirus in Latin America } (Aug 2020),
  \url{https://www.as-coa.org/articles/coronavirus-latin-america}

\bibitem{first_case_3}
Braziliense, C.: { Casos suspeitos de coronavírus são registrados em Porto
  Alegre e Curitiba } (Jan 2020),
  \url{https://www.correiobraziliense.com.br/app/noticia/brasil/2020/01/28/interna-brasil,823972/casos-suspeitos-de-coronavirus-sao-registrados-em-porto-alegre-e-curit.shtml}

\bibitem{paper3}
Chire~Saire, J.E.: How was the mental health of colombian people on march
  during pandemics covid19? medRxiv  (2020). \doi{10.1101/2020.07.02.20145425},
  \url{https://www.medrxiv.org/content/early/2020/07/04/2020.07.02.20145425}

\bibitem{paper5}
Chire~Saire, J.E.: Infoveillance based on social sensors to analyze the impact
  of covid19 in south american population. medRxiv  (2020).
  \doi{10.1101/2020.04.06.20055749},
  \url{https://www.medrxiv.org/content/early/2020/04/11/2020.04.06.20055749}

\bibitem{paper1}
Chire~Saire, J.E., Oblitas, J.: Covid19 surveillance in peru on april using
  text mining. medRxiv  (2020). \doi{10.1101/2020.05.24.20112193},
  \url{https://www.medrxiv.org/content/early/2020/05/25/2020.05.24.20112193}

\bibitem{paper2}
Chire~Saire, J.E., Pineda-Briseno, A.: Text mining approach to analyze
  coronavirus impact: Mexico city as case of study. medRxiv  (2020).
  \doi{10.1101/2020.05.07.20094466},
  \url{https://www.medrxiv.org/content/early/2020/05/12/2020.05.07.20094466}

\bibitem{paper4}
Drias, H.H., Drias, Y.: Mining twitter data on covid-19 for sentiment analysis
  and frequent patterns discovery. medRxiv  (2020).
  \doi{10.1101/2020.05.08.20090464},
  \url{https://www.medrxiv.org/content/early/2020/05/18/2020.05.08.20090464}

\bibitem{first_case_4}
Folha: { Brasil confirma primeiro caso do novo coronavírus } (Jan 2020),
  \url{https://www1.folha.uol.com.br/equilibrioesaude/2020/02/brasil-confirma-primeiro-caso-do-novo-coronavirus.shtml}

\bibitem{peak_cases}
Globo: { Brasil tem 13.993 mortes e 202.918 casos confirmados de novo
  coronavírus, diz ministério } (May 2020),
  \url{https://g1.globo.com/bemestar/coronavirus/noticia/2020/05/14/brasil-tem-13993-mortes-causadas-pelo-novo-coronavirus-diz-ministerio.ghtml}

\bibitem{first_case_2}
Globo: { Ministério investiga caso suspeito de coronavírus em MG e pede que
  viagens à China sejam evitadas } (Jan 2020),
  \url{https://g1.globo.com/ciencia-e-saude/noticia/2020/01/28/ministerio-da-saude-confirma-caso-suspeito-de-coronavirus-em-mg.ghtml}

\bibitem{paper9}
Ji, X., Tang, Z., Wang, K., Li, X., Li, H.: Analysis of epidemic situation of
  new coronavirus infection at home and abroad based on rescaled range (r/s)
  method. medRxiv  (2020). \doi{10.1101/2020.03.15.20036756},
  \url{https://www.medrxiv.org/content/early/2020/03/20/2020.03.15.20036756}

\bibitem{covid_dataset}
Mello, L.E., Suman, A., Medeiros, C.B., Prado, C.A., Rizzatti, E.G., Nunes,
  F.L.S., Barnabé, G.F., Ferreira, J.E., Sá, J., Reis, L.F.L., Rizzo, L.V.,
  Sarno, L., de~Lamonica, R., Maciel, R.M.d.B., Cesar-Jr, R.M., Carvalho, R.:
  {Opening Brazilian COVID-19 patient data to support world research on
  pandemics} (Jul 2020). \doi{10.5281/zenodo.3966427},
  \url{https://doi.org/10.5281/zenodo.3966427}

\bibitem{paper8}
Mizumoto, K., Kagaya, K., Chowell, G.: Early epidemiological assessment of the
  transmission potential and virulence of coronavirus disease 2019 (covid-19)
  in wuhan city: China, january-february, 2020. medRxiv  (2020).
  \doi{10.1101/2020.02.12.20022434},
  \url{https://www.medrxiv.org/content/early/2020/06/15/2020.02.12.20022434}

\bibitem{crisp2000}
Shearer, C.: The crisp-dm model: The new blueprint for data mining. Journal of
  Data Warehousing  \textbf{5}(4) (2000)

\bibitem{paperbr_1}
{Waheed}, A., {Goyal}, M., {Gupta}, D., {Khanna}, A., {Al-Turjman}, F.,
  {Pinheiro}, P.R.: Covidgan: Data augmentation using auxiliary classifier gan
  for improved covid-19 detection. IEEE Access  \textbf{8},  91916--91923
  (2020)

\bibitem{paper10}
Yuan, X., Hu, K., Xu, J., Zhang, X., Bao, W., Lynch, C.F., Zhang, L.: State
  heterogeneity of human mobility and covid-19 epidemics in the european union.
  medRxiv  (2020). \doi{10.1101/2020.06.10.20127530},
  \url{https://www.medrxiv.org/content/early/2020/06/12/2020.06.10.20127530}

\end{thebibliography}

\end{document}